# Grain boundary scars on spherical crystals

*Thomas Einert[1*], Peter Lipowsky[1*], Jörg Schilling[1], Mark J. Bowick[2] and Andreas R. Bausch[1]*

[1]Lehrstuhl für Biophysik E22, TU München, 85747 Garching, Germany

[2]Physics Department, Syracuse University, Syracuse NY 13244-1130, USA

\* contributed equally to the work

We present an experimental system suitable for producing spherical crystals and for observing the distribution of lattice defects (disclinations and dislocations) on a significant fraction (50%) of the sphere. The introduction of fluorescently labeled particles enables us to determine the location and orientation of grain boundary scars. We find that the total number of scars and the number of excess dislocations per scar agree with theoretical predictions and that the geometrical centers of the scars are roughly positioned at the vertices of an icosahedron.

When crystals form on curved surfaces it is known that new defect structures may arise even at zero temperature. Dislocations (consisting of pairs of tightly bound 5 and 7 fold coordinated defects) form in the ground state of sufficiently large and rigid curved crystals because they lower the total elastic energy [1-5]. In the specific case of spherical crystals these dislocations may be viewed as relieving the elastic strain of the 12 isolated disclination defects (5-fold coordinated particles) necessarily present given the topology of the sphere. They are present, above a critical particle number of order 400, in the form of novel freely terminating high-angle grain boundaries dubbed scars [6].

Recently, these grain boundary scars were observed experimentally by bright field microscopy visualizations of polystyrene-coated water droplets dispersed in oil. The images analyzed were of spherical caps subtending a solid angle ranging from 10 to 20% of the full $4\pi$ solid angle of the sphere. Since one expects approximate icosahedral symmetry for the ground state it is actually sufficient to analyze 1/12 of the surface of the sphere to test the predictions of equilibrium statistical mechanics. We were able to confirm theoretical estimates for the number of excess dislocations per scar and also to analyze the dynamics of both particles and dislocation defects [7]. In order, however, to properly check the global features of the crystalline ground state it is essential to observe the location and structure of several, ideally *all*, scars on the sphere – thus a bigger area has to be imaged. This can be accomplished by using laser scanning confocal microscopy (LSCM) to determine the three dimensional (3D) coordinates of every colloidal bead. For triangulation it is also necessary that the optical resolution and contrast of the image be sufficient for automatic bead detection.

In this paper we introduce an experimental system for observing the 3D positions of defects on spherical surfaces of radius R. We have developed a method for automatically tracking particles and simultaneously Delaunay triangulating the lattice they form – in this way each particle configuration may be paired with the resultant lattice defect array on the sphere. We studied crystal-coated water droplets with dimensionless system size $R/a$ between 11 and 14, where $a$ is the mean particle spacing. This range of sizes corresponds to an ordered coating of the entire sphere by between 1700 and 2800 beads. The scars observed typically contained between 4 and 6 excess dislocations and were found to be centered, on average, at the vertices of an icosahedron. The observed displacements from broad icosahedral symmetry are most likely due to the combined effects of thermal fluctuations and local stresses resulting from additional "impurity" clusters of beads not equilibrated with the crystal. Our results confirm that the grain boundary scars themselves are *inherent* structural features of the *ground state* of sufficiently large spherical crystals, rather than extraneous defect arrays arising from thermal fluctuations or external stresses.

Our model system for studying curved two dimensional (2D) crystals is a Pickering emulsion[8, 9]; specifically a liquid-liquid emulsion of water droplets in toluene stabilized by fluorescently labeled silica microspheres 1.5 µm in diameter[10, 11]. To prevent particle aggregation the microspheres were silanized with n-octyltrimethoxysilane (ABCR, Karlsruhe, Germany). The increased hydrophobicity was necessary to obtain stable and regular structures at the oil-water interface [12]. In order to minimize the water content of the final dispersion, fluorescently labeled silica particles (Sicastar Green F, Micromod, Rostock, Germany) were repeatedly dried and redispersed, first in ethanol, then in acetone, before they were transferred to anhydrous toluene. 300 µl of silane was added to the final volume of 5 ml together with a catalyser (25 µl butylamine). After incubation in an ultrasonic bath for one hour, under continuous stirring, the particles were repeatedly centrifuged and washed with acetone and re-dispersed in ethanol, where they were stored at a final concentration of 10 mg/ml. Finally the particles were centrifuged again and transferred into 10 ml of toluene. For the aqueous phase of the emulsion, 300 µl of double distilled water was added to the toluene immediately before optical inspection. The emulsion was formed by gentle mixing with a pipette.

Silanized silica microspheres form a triangular lattice at the oil-water interface at lower areal densities than polystyrene beads, probably due to the presence of long range interactions [12]. The lattice spacing of these dilute crystals is about five to seven times the particle diameter. Adsorption of additional particles at the oil-water-interface increases the areal density of particles and a dense crystalline state is eventually formed with a lattice spacing of about 1.2 particle diameters. At this stage the mobility of the particles is restricted. The crystalline structure is therefore quite stable over the very long scanning times (10 min) of the confocal imaging system (LSM510, Zeiss, Germany). The low emission intensity of the particles necessitated these relatively long scanning times in order to obtain a signal-to-noise ratio sufficiently high for automated evaluation of the micrographs. The initial high mobility of the particles during the formation of the crystal leads to equilibrium configurations. Images were taken with a 63x oil objective (numerical aperture 1.4). The fluorescence excitation was achieved with a laser of 488nm and emission was observed with the help of a 505nm long pass filter.

Due to the different refractive indices of water ($n_w$=1.33) and toluene ($n_t$=1.49) the water drop acts like a downscaling lens: the hemisphere which is farther away from the microscope's objective appears to be smaller than the one closer to it. Ostwald ripening of the emulsion was observable: larger crystals grew at the expense of smaller crystals. As small crystals shrink, with a fixed number of bound silanized silica beads on the surface, they eventually buckle or indent. Irregularly buckled crystal structures were observable after several hours and were excluded from further analysis.

Three dimensional coordinates for each colloidal particle were determined by a correlation algorithm. A test bead - the correlation kernel - is compared with a region in the original 3D image and a correlation image is produced:

$$Correlation(\vec{R}) = \sum_{\vec{r} \in Kernel} Image(\vec{R}+\vec{r}) \cdot Kernel(\vec{r}),$$

where Correlation, Image and Kernel denote the 3D arrays of the intensities of the three images respectively and $\vec{R}$ and $\vec{r}$ are vectors with three components and simply point to a pixel in one of the images. The correlation algorithm results in higher intensities the better the region around the corresponding pixel in the original image data resembles the kernel. The bead positions are thus determined by the local maxima in the correlation image. This procedure worked extremely well, even for the far hemisphere shrunk by lensing, and failed only in the equatorial region or regions with very low contrast. We were able to omit the calculation of a normalization constant to save computational time, as almost no overly bright regions were observable. As a further check on the algorithm we fed the positions of the beads, as determined by the correlation algorithm, back into the original image data; in this way tracking artifacts could be corrected manually.

To eliminate the lens effect of the water drop we fitted a sphere to the larger undistorted hemisphere. The coordinates of all the beads in the smaller hemisphere were then collectively shifted and rescaled, with respect to the sphere's centre, so that the deviation from the fitted sphere was minimized. This correction for optical distortion is essential for a proper triangulation of the crystal and also yielded the most accurate determination of the sphere radius R and the location of its centre. The bead positions obtained this way are then mapped to a Delaunay triangulation of the sphere. For a flat 2D lattice the Delaunay tessellation consists of all triangles whose circumcircles contain only the three vertices of that triangle. This may be adapted to the case of the two-sphere by constructing the set of triangles on the sphere's surface whose cones (defined by the circumcircle of a triangle and its apex at the centre of the sphere) enclose only the three vertices of the associated triangle. We implemented a growth algorithm to build the tessellation. One first assumes a region of the crystal is already triangulated and then chooses an edge which belongs to only one triangle. Such an edge is necessarily on the border between the currently triangulated region and a non-triangulated region. Then one finds a vertex from the set of all lattice sites so that the triangle formed by that vertex and the edge satisfies the defining cone condition above. This triangle is added to the set of existing triangles and the process iterated. A seed edge for the algorithm may be found by joining an arbitrary vertex to its nearest neighbor. The spherical Delaunay algorithm was implemented with MatLab. Given a triangulation one has a complete identification of all the topological defects on the lattice. Linear arrays of dislocations, separated by less than 2 edges, were identified as scars.

The colloidal beads were observed to self-assemble into spherical crystals, as displayed in Fig. 1. A few aggregates inevitably form in the oil phase. Although some of these aggregates stuck to the spherical crystals, or incorporated themselves into the crystal structure, only local disturbances of the triangular lattice resulted. We choose to study only crystals with system size $R/a$ in the range 11 to 14. For this range of sizes it is expected, theoretically, that the ground state will contain grain boundary scars with 3-5 excess dislocations [6, 13]. The spherical crystal displayed in Fig. 1 has a radius of 22 µm and a mean bead spacing of 1.8 µm; the system size is therefore R/a≈12. Most of the crystalline surface forms an excellent triangular lattice, as shown in Fig. 2. In the equatorial area, however, the optical resolution was insufficient to determine particle positions accurately and the resultant triangulation was not reliable there. We were able to triangulate a solid angle of more than $\pi$ in each hemisphere and thus more than 50% of the entire sphere is triangulated. This is more than sufficient to address the nature of the global distribution of disclination defects and is a dramatic improvement over the 10 to 20% of our previous work. The triangulations reveal prominent scars and a few isolated dislocations distributed in a regular (non-defective) background lattice. For the crystal shown in Fig. 2, there are 6 clear grain boundary scars. This is exactly what one would expect from observing 50% of the sphere since there should be 12 scars in all. Each scar has one excess 5 disclination, as required, and contains between 3 and 6 elastically bound dislocations. The average number of excess dislocations (4) is in excellent agreement with theoretical predictions [6, 13]. We analyzed 4 crystals in detail and found qualitatively and quantitatively similar results in each case. We do not yet have sufficient data to provide a meaningful statistical analysis of the number of excess dislocations per scar. The scars are noticeably curved (see Fig. 2). At finite temperature additional configurations are available with curved scars and thus the entropic gain in the free energy can explain the tendency of scars to bend.

The angular span of a scar is also a universal number independent of the details of the microscopic potential. Theory [13] predicts that scars should span $2\cos^{-1}(5/6) = 67^0$. The longest scar we found for the crystal shown spans $64^0$, with several others of similar length. Several shorter scars have dislocations a few lattice spacings away that would probably be bound to the scar at zero temperature.

Some of the straighter scars were used to further analyze the nature of the grain boundary itself. The angular mismatch of crystallographic axes across a grain boundary scar is found to be continuously varying, as shown in Fig. 3. It varies from approximately 30° at the centre of a scar to zero at the end. This change of the angle results from the curvature of the surface: the further one is away from the centre of the scar the more the elastic distortion of the lattice is screened by the curvature of the sphere and the interior dislocations, resulting in less angular mismatch between the orientation of crystallographic axes on each side of the grain boundary. Eventually, the total disclination charge is effectively screened completely and the grain boundary freely terminates. This phenomenon is energetically forbidden in flat space [1, 13] and is another signature effect of spatial curvature revealed in our experiment. The particular angles we observe are undoubtedly affected by thermal fluctuations of the grain boundary as well [7]. Numerical simulations of particles distributed on the sphere and interacting with repulsive power law potentials indicate that this effect is more regular at zero temperature [14].

To examine the alignment of the scars with respect to each other we measured the angles between the scars. As the reference point for each scar we chose the 5-fold coordinated bead in the middle of the scar. If each scar is centered at the vertex of an icosahedron, as theory predicts, we would expect four distinct angles to occur: $0°$, $63.4°$, $116.6°$ and $180°$: these angles follow from the classical geometry of convex polyhedra. Some of the observed angles were close to one of these four values but not all. Our analysis here may be visualized by fitting an icosahedron to the 5-fold coordinated beads in the middle of each scar (as shown in Fig. 4). While some scars are nearly centered on an icosahedral vertex, others are not. Nor did we find any clear trend in the relative orientation of the scars. Both these latter findings can be attributed to the marked effect of thermal fluctuations on the colloidal configurations and the complexity of the energy landscape for interacting particles on the sphere which can lead to almost degenerate ground states with different defect arrangements. The presence of adherent clusters at some locations on the crystal could also perturb the quasi-long-range translational order characteristic of 2D crystals, despite the fact that they are clearly not disturbing the short range order of the crystal. It is possible that the elastic interaction energy between distinct scars is influenced by these non-topological "impurities" and that they are consequently not perfectly equilibrated globally. This speculation is also consistent with the finding that the scars on the hemisphere with no adherent clusters are centered close to the vertices of an icosahedron, while this fails to be true on the other hemisphere with adherent clusters.

The silica-bead coated water droplets studied here have allowed us to observe, for the first time, the global spatial distribution of grain boundary scars in a spherical crystal. Our findings are consistent with the necessary topological constraints associated with spherical topology – extrapolating to the whole surface we would have 12 grain boundary scars total. In contrast to the curved crystal systems previously studied the equilibrated dislocations we observe here are not very mobile over the time scale of the experiment [7]. Although the structural features are consistent with the broad predictions of equilibrium statistical mechanics there is most likely some kinetic trapping occurring. Nevertheless we confirm both the presence of the correct number of scars. This demonstrates the *intrinsic* nature of grain boundary scars in sphereland. We have found that experimentally inevitable impurity clusters only marginally disturb the crystalline order. Nevertheless there is some departure from global icosahedral symmetry – this may be due to significant thermal fluctuations or to cluster impurities and is under further investigation. One can speculate that melting in this system would be initiated at the grain boundary scars and thus the presence of scars may be crucial in understanding phase transitions in spherical crystal structures [15, 16].

## Acknowledgements

This work was supported by the DFG (BA2029-5) and the Fonds der Chemischen Industrie. The work of MJB was supported by the National Science Foundation through Grant No. DMR-0219292 (ITR).

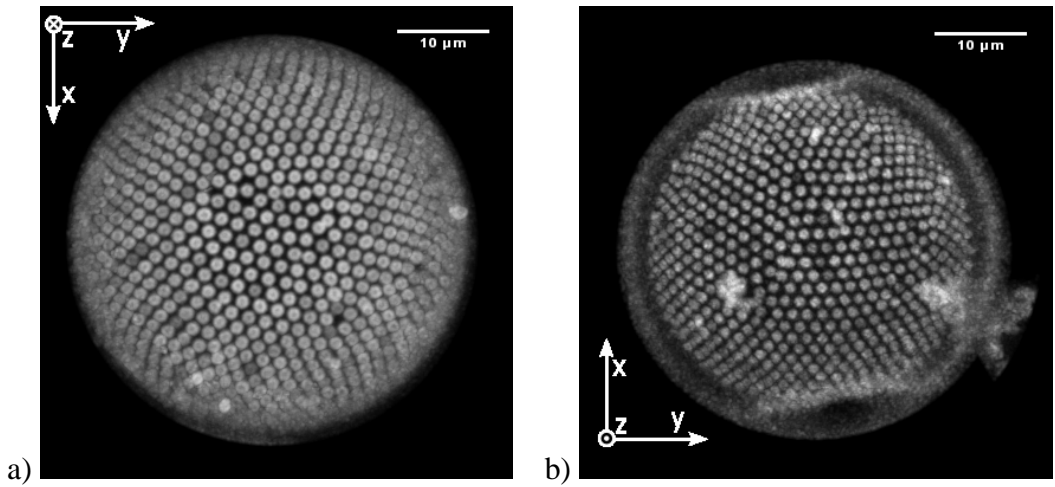

Fig.1: Projection of a confocal image of a spherical crystal (radius R=22 µm and dimensionless system size R/a = 12): a) the lower hemisphere with the optical axis pointing into the image; b) the upper hemisphere as viewed from above with the optical axis pointing out of the image. The colloids which were not adsorbed at the interface are not shown in the image. The scale bar is 10µm. The seemingly irregular shape of the crystal does not arise from Ostwald ripening induced buckling but rather from the optical distortion caused by the refraction of light at adjacent water droplets not shown in the images.

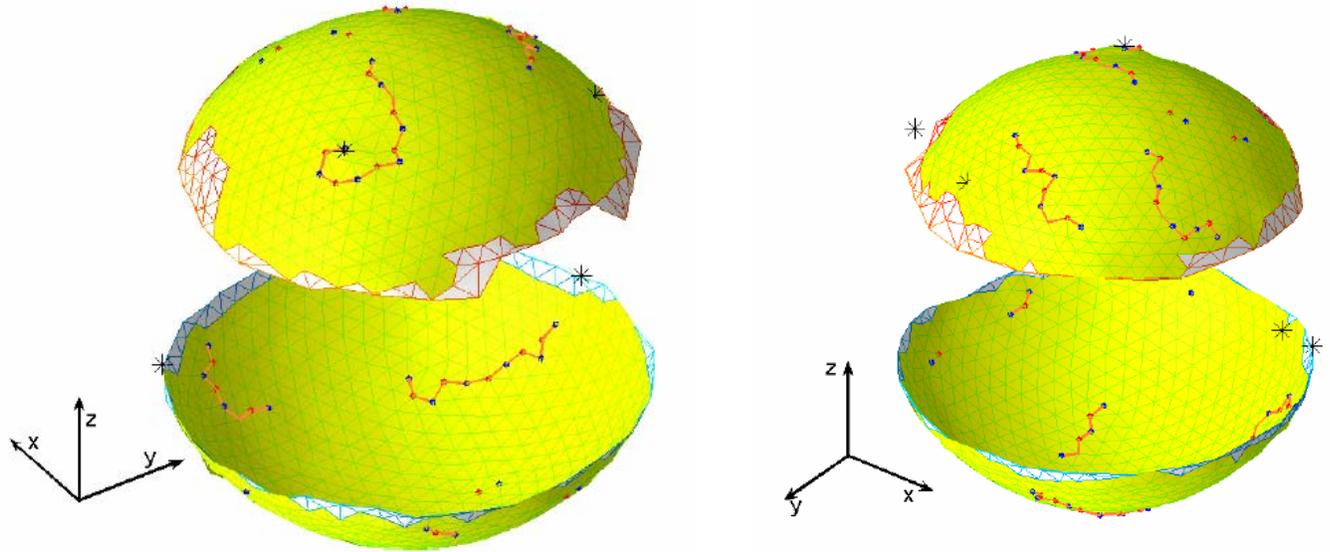

**Fig. 2:** a) The spherical Delaunay triangulation of the droplet displayed in Fig. 1 (R = 22µm, R/a = 12, N ≈ 2150);  b) The same sphere rotated 180° around the optical axis. The blue and red marked lattice sites denote five-fold and seven-fold coordinated beads respectively. The asterisks indicate the positions of non-invasive adherent clusters. The observed scars contain between 3 and 6 dislocations plus an excess five-fold disclination. Scars too close to the equatorial region cannot be reliably analyzed. A few isolated dislocations, very likely thermal excitations, are also visible. The scale is provided by the 10µm coordinate axes in the lower left.

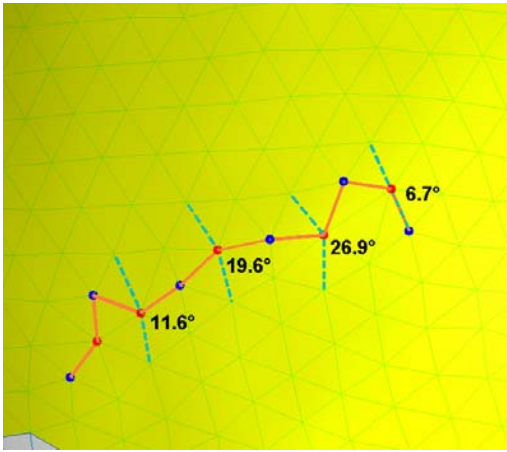

**Fig. 3:** The rotation angle between crystallographic each side of a grain boundary scar varies continuously along the scar, as shown in the magnified view of the scar located on the right hand side of the bottom hemisphere of Fig. 2 a.

a)

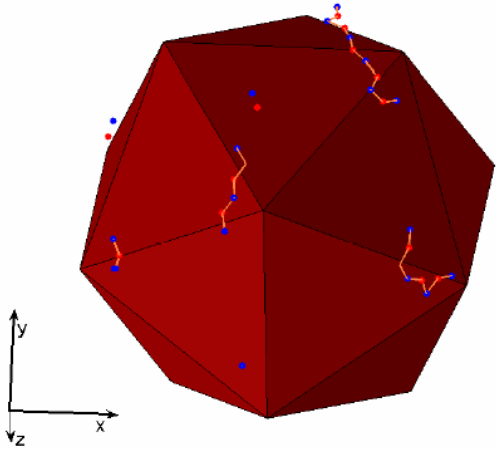 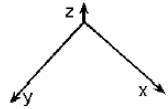 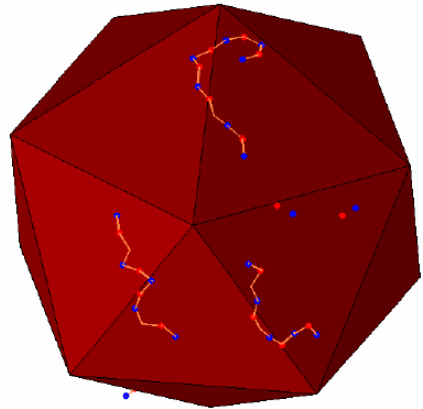

b)

**Fig. 4:** Global alignment of grain boundary scars. The experimentally determined scars were plotted on an icosahedron, with the orientation of the icosahedron chosen to maximize the overlap between the scar centers and the icosahedral vertices. The viewpoints in a) and b) are similar to the projections shown in Fig. 1. Note that the scars do not lie exactly on the vertices. Deviations from icosahedral symmetry are particularly evident in the upper half shown of b). In this hemisphere 3 colloidal clusters were adhering to the crystal. The deviations from icosahedral symmetry are probably due to a combination of thermal fluctuations and local disturbances by adhering "impurity" clusters. The scale is provided by the 10µm coordinate axes shown.

# For Table of Contents Use Only

Grain boundary scars on spherical crystals
Thomas Einert1*, Peter Lipowsky1*, Jörg Schilling1, Mark J. Bowick2 and Andreas R. Bausch1

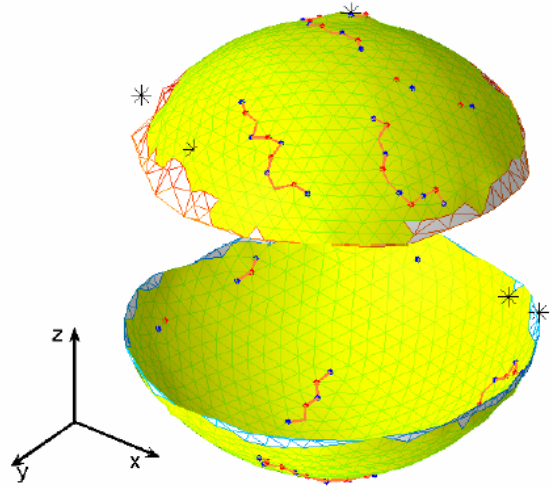